\documentstyle[epsfig,longtable]{aipproc}

\begin{document}
\title{The CANGAROO-III Project}

\renewcommand{\thefootnote}{\fnsymbol{footnote}}
\author{Masaki Mori$^1$\footnote{%
E-mail: {\tt morim@icrr.u-tokyo.ac.jp}},
S.A.~Dazeley$^2$,
P.G.~Edwards$^3$,
S.~Gunji$^4$, S.~Hara$^5$,
T.~Hara$^6$, J.~Jinbo$^7$,
A.~Kawachi$^1$, T.~Kifune$^1$,
H.~Kubo$^5$,
J.~Kushida$^5$, Y.~Matsubara$^8$,
Y.~Mizumoto$^9$,
M.~Moriya$^5$,
H.~Muraishi$^{10}$, Y.~Muraki$^8$,
T.~Naito$^6$, K.~Nishijima$^7$,
J.R.~Patterson$^2$, M.D.~Roberts$^1$,
G.P.~Rowell$^1$, T.~Sako$^{8,11}$,
K.~Sakurazawa$^5$, Y.~Sato$^1$, R.~Susukita$^{12}$,
T.~Tamura$^{13}$,
T.~Tanimori$^5$, S.~Yanagita$^{10}$,
T.~Yoshida$^{10}$, T.~Yoshikoshi$^1$, and
A.~Yuki$^8$  }
\address{$^1$Institute for Cosmic Ray Research, University of Tokyo
Tanashi, Tokyo 188-8502, Japan\\
$^{2}$Department of Physics and Mathematical Physics, University of
   Adelaide, South Australia 5005, Australia\\
$^{3}$Institute of Space and Astronautical Science,
   Sagamihara, Kanagawa 229-8510, Japan\\
$^{4}$Department of Physics, Yamagata University,
Yamagata 990-8560, Japan\\
$^{5}$Department of Physics, Tokyo Institute of Technology,
        Meguro, Tokyo 152-8551, Japan\\
$^{6}$Faculty of Management Information, Yamanashi Gakuin Univeristy,  Kofu,
Yamanashi 400-8575, Japan\\
$^{7}$Department of Physics, Tokai University,
 Hiratsuka, Kanagawa 259-1292, Japan\\
$^{8}$STE Laboratory, Nagoya University,
   Nagoya, Aichi 464-8602, Japan\\
$^{9}$National Astronomical Observatory, Tokyo 181-8588, Japan\\
$^{10}$Faculty of Science, Ibaraki University,
   Mito, Ibaraki 310-8521, Japan\\
$^{11}$LPNHE, Ecole Polytechnique. Palaiseau CEDEX 91128, France\\
$^{12}$Computational Science Laboratory, Institute of Physical and Chemical
   Research, Wako, Saitama 351-0198, Japan\\
$^{13}$Faculty of Engineering, Kanagawa University,
 Yokohama, Kanagawa 221-8686, Japan
}

\maketitle

\begin{abstract}
The CANGAROO-III project, which consists of an array of four
10~m imaging Cherenkov telescopes, has just started being constructed
in Woomera, South Australia, in a collaboration between Australia and Japan.
The first stereoscopic observation of celestial high-energy gamma-rays
in the 100~GeV region with two telescopes will start in 2002, and the four
telescope array will be completed in 2004.
The concept of the project and the expected performance are discussed.
\end{abstract}

\section*{Introduction}

Following the CANGAROO-I (3.8~m) and
CANGAROO-II (7~m) telescopes,
CANGAROO-III is a project to study celestial gamma-rays
in the 100 GeV region utilizing a stereoscopic observation
of Cherenkov light flashes with an array of four 10-meter
telescopes.
The CANGAROO-II telescope (hereafter C-II), which has a 7-meter
reflector and has been operational since 1999 May
\cite{Tanimori99} \cite{Mori99} \cite{Kubo99}, is going to be
expanded in early 2000 by adding more small mirrors and will be the
first 10~m telescope of this array.

\begin{figure}[b!]
\centerline{\epsfig{file=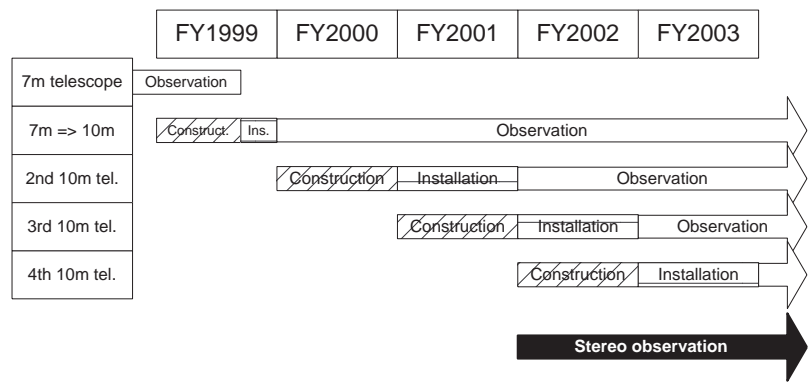,height=5cm}}
\vspace{10pt}
\caption{Schedule of the CANGAROO-III project. Note that the
Japanese fiscal year is from April to March.}
\label{fig:sched}
\end{figure}

The CANGAROO-III project started in April 1999 and is planned as a five-year
program. The schedule is shown in Figure \ref{fig:sched}.
 This year we will expand the
7~m telescope to 10~m, and the second year we will build the
second telescope which will be installed in the third year.
The other two telescopes will be installed in the fourth and fifth
years. 
Each telescope will be set on a corner of a diamond of about 100~m side
in order to have a maximum number of pairs of telescopes of the
same baseline length.
The first stereoscopic observation will be performed in 2002 and the full
four telescope will be in operation in 2004.

\section*{Expansion of CANGAROO-II}

\begin{figure}[b!]
\centerline{\epsfig{file=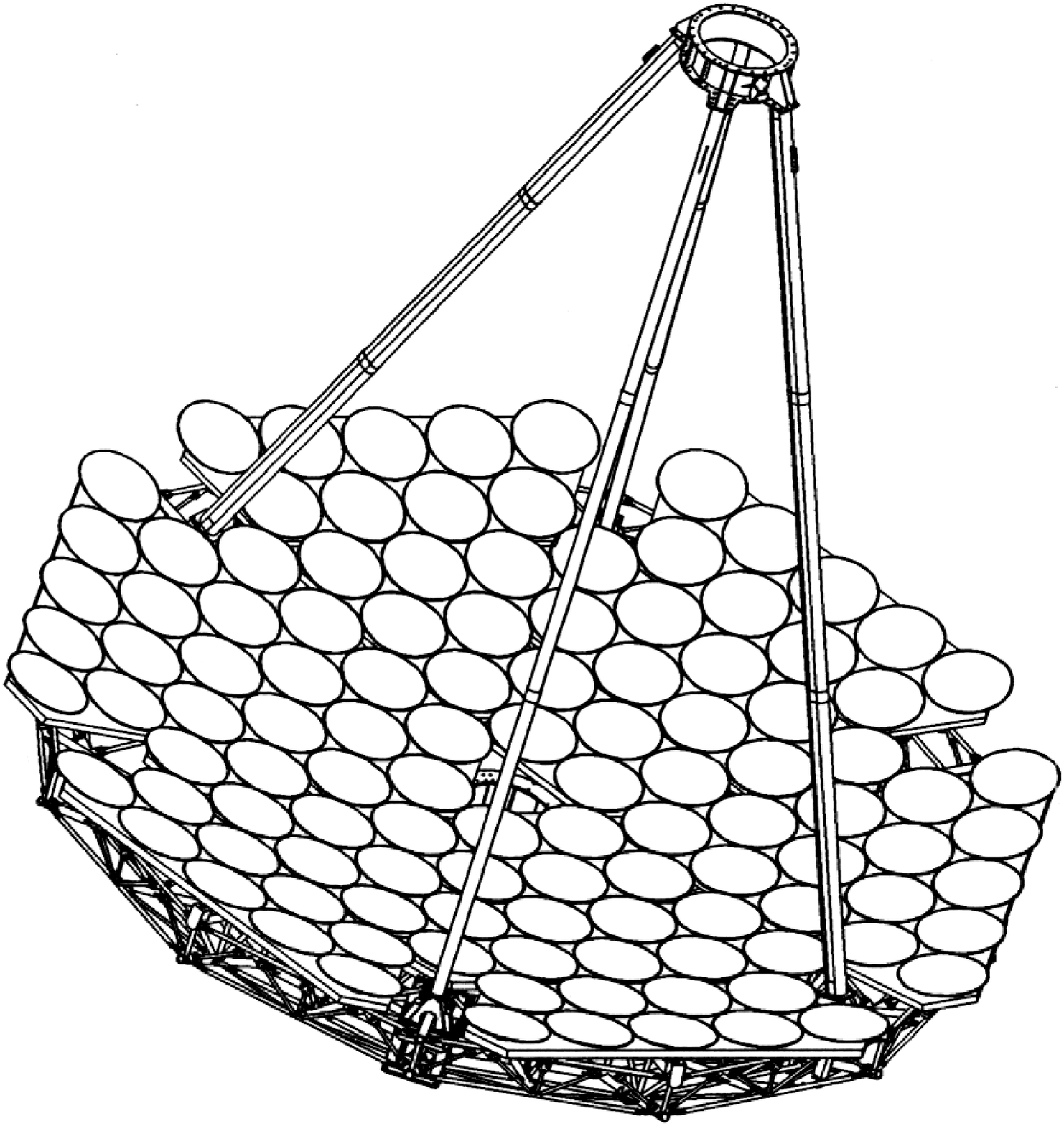,height=6cm}}
\vspace{10pt}
\caption{A sketch of the 10~m reflector.}
\label{fig:10m}
\end{figure}

Expansion of the 7~m telescope to 10~m is simple.
Since C-II is originally designed as a 10~m telescope, all we have
to do is add 54 mirrors and tune their attitude (Figure \ref{fig:10m}).
This work will be completed in early 2000.

Additional outer mirrors will worsen the point image
at the focus but it is not a serious problem. Simulations
show that the concentration of photons in one pixel
will be reduced from 56\% to 42\% at the center, and from
50\% to 36\% at one-degree off-axis. In any case, the
number of collected photons will be almost doubled,
reducing the energy threshold by a factor of two.

\section*{CANGAROO-III Telescope Design}

At this stage we will use basically the same design for
the support structure and the driving mechanism as the C-II telescope,
which is originally designed as a 10-meter telescope
and has been proved to work well.
The reflector will have a parabolic, composite
mirror consisting of 114 small mirrors of 80~cm in diameter
\cite{Kawachi99}.
The focal length will be 8~m if we use the same mirrors.

Mirrors made of plastic laminates used for C-II
are very light and pose little stress on the support
structure.
Observation of star images at various zenith angles
showed the deformation of mirrors was negligible.
But the image quality of these mirrors are not as good as glass-made mirrors
since they are made by molding: thus we are still investigating
other possibilities.
The attitude of each mirror will be controlled by stepping motors
as for the present 7~m telescope.
Tuning this number of small mirrors to a common focus is not
a simple task. For C-II we tuned the mirrors
one by one using lids to cover the other mirrors,
but this is not easy for larger numbers of mirror segments.

The prime focus camera will be similar to the present
CANGAROO-II camera consisting of 512 half-inch photomultipliers and
subtending about 3 degrees, but the optimization
for stereoscopic observation is underway.

The electronics and data acquisition system will be improved
to match higher data rates.
In any case, we take timing information of each signal, in addition
to pulse height, to utilize the isochronous nature of our 
parabolic reflector.
For stereoscopic observation, we must introduce an inter-telescope
trigger to compensate for geometrical delays using programmable delays
between telescopes.
The local triggers will be as frequent as 1 kHz but the
delayed coincidences at the main trigger
will be reduced to about 100 Hz, we hope.

\section*{Stereo Simulation}

Here we briefly show some results of simulations of stereo observations
 \cite{Hara99}. This work
was done before the whole CANGAROO-III project was approved
and takes only two telescopes into account, but the result is valid
if we use a twofold coincidence in the inter-telescope trigger.

\begin{figure}[b!]
\centerline{\epsfig{file=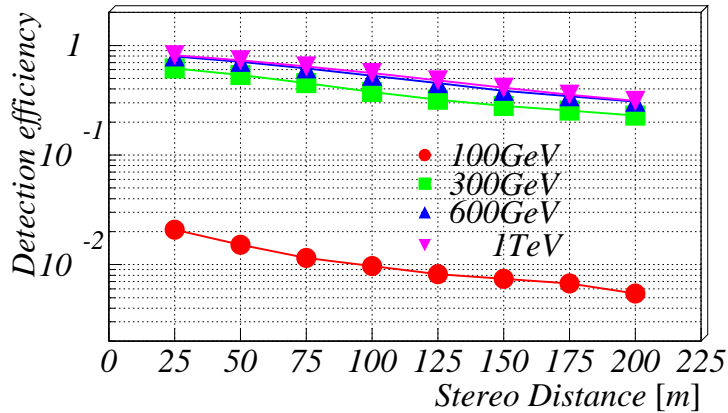,height=7cm}}
\vspace{10pt}
\caption{Detection efficiency as a function of the baseline separation
of two telescopes. Here we define the efficiency as a fraction of
triggered gamma-rays when we simulated gamma-rays going vertically
and having cores in a circle of 180~m radius.}
\label{fig:eff}
\end{figure}

\begin{figure}[b!]
\centerline{\epsfig{file=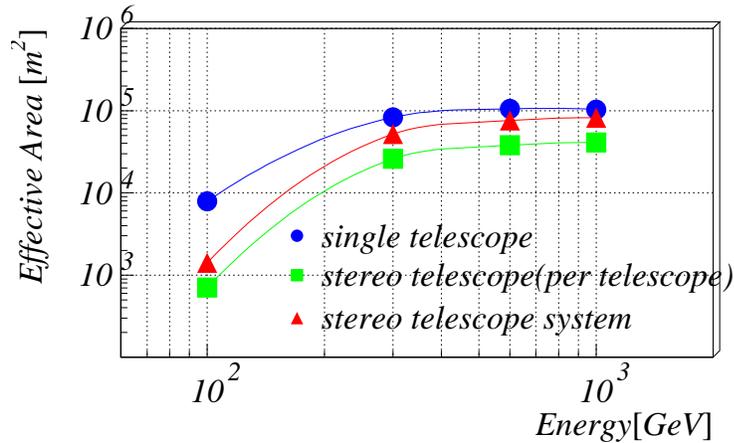,height=7cm}}
\vspace{10pt}
\caption{Detection area as a function of gamma-ray enegy for
single and stereo observations.}
\label{fig:monoste}
\end{figure}

The detection efficiency as a function of baseline length between telescopes
is given in Figure \ref{fig:eff}. If we cut some detected events
using the core distance, the energy resolution will be better
and an angular resolution less than $0.1^\circ$ can be achieved if we use
the baseline longer than 100~m. Thus we will adopt the
baseline length of around 100~m, which agrees with other
calculations.
Figure \ref{fig:monoste} is a comparison of effective area and energy
resolution between single and stereo observations.

\section*{Expected performance}

\begin{figure}[b!]
\centerline{\epsfig{file=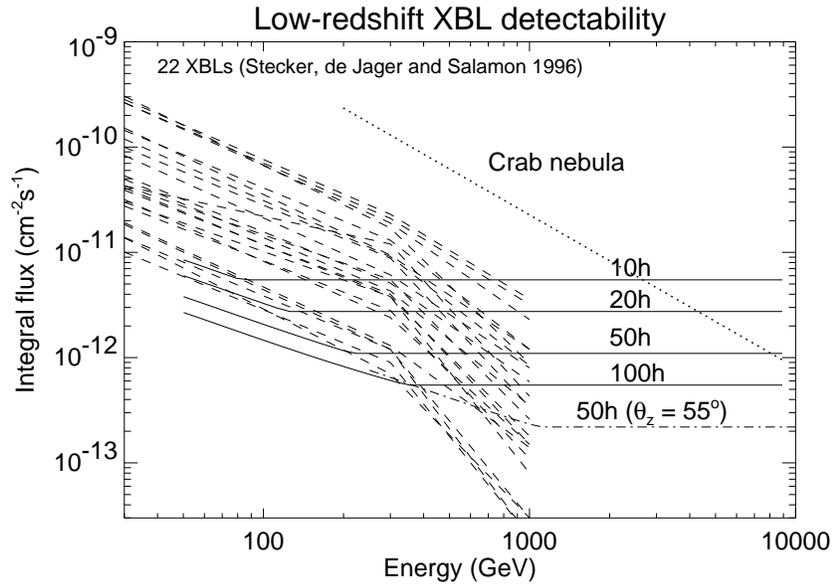,height=8cm}}
\vspace{10pt}
\caption{Sensitivity of the 10~m telescope and predicted fluxes from X-ray
selected BL Lacs (dashed lines) \protect\cite{Stecker96}.
Here ``10h'' means 10 hours of observation and so on.}
\label{fig:c3sens}
\end{figure}

Expected sensitivity assuming one 10~m telescope shown in Figure
\ref{fig:c3sens} is around $10^{-12}$~cm$^{-2}$s$^{-1}$
above threshold energy of $100\sim200$~GeV, and
we may detect many EGRET sources in tens of hours
of observation if their spectra extend to higher energies.
Also shown are gamma-ray spectra of 22 X-ray selected BL Lacs which
are predicted by Stecker et al.\ \cite{Stecker96}, however only 5 are
in the southern hemisphere.
We note that observations at other wavelengths, especially
ground-based ones, are rather biased to the northern hemisphere sky
and there are undoubtedly more candidate XBLs in the southern sky.

\section*{Observation targets}

Table \ref{tab:list} shows the list of objects observed by the
CANGAROO 3.8m telescope for its 6 years of operation.
We had been given preference to Galactic sources because of
the rather high threshold energy ($\sim2$ TeV) of the 3.8m
telescope, but we may spend more time on extragalactic objects
taking account of the lower threshold of new telescopes.
One can see from the table 
we have needed more than 50 hours of observation
at least as the necessary condition to conclude ``positive
detection" with sufficient statistics on the number of
gamma-rays. In addition, the imaging Cherenkov technique
still suffers from systematic errors which are not negligibly
small when compared with gamma-ray signal strength from even
``strong sources", and careful estimation on the experimental 
errors is indispensable by using the data spanning over 
a long period of observation.
We performed survey observations of shorter duration on many 
sources, which possibly provide a chance of time varying activities
of episodic flares, as well as the objects like X-ray binaries
which might be ``strong sources" if the claims in earlier days
are true.
The prime efforts of CANGAROO-III will be on those types
of objects appearing as top-ranked sources in the table,
extending a systematic survey on more sources.
In the case of sources of soft spectra, better statistics
in the 100~GeV energy region will enable us to detect them 
in 10 to 20 hours of observation.
However, we still have shortage of total observation time 
available during a year.
It is necessary to develop world-wide efforts for more
new types of high-energy gamma-ray sources in collaboration 
with other groups proposing next-generation telescopes.

\begin{table}[b!]
\caption{A list of objects observed by the CANGAROO 3.8m telescope
from July 1992 to September 1998 for more than 15 hours in the order
of observation time, including bad weather runs.
Total observation time is about 2,000 hours.
Off-source runs are roughly the same duration but are not listed here. }
\label{tab:list}
\begin{tabular}{ldll}
Object & \multicolumn{1}{c}{Observation} & Remark & Reference\\
       & \multicolumn{1}{c}{time (hr)} &        & \\
\tableline
PSR1706$-$44     &308.5 &Plerion & \cite{Kifune95}\\
Vela             &252.6 &Plerion & \cite{Yoshikoshi97}\\
Crab             &193.8 &Plerion & \cite{Tanimori94} \cite{Tanimori98}\\
PSR1259$-$63     &167.4 &Pulsar binary  & \cite{Sako97}\\
PSR1509$-$58     &161.5 &Plerion & \cite{Sako97} \cite{Sako99}\\
W28              &121.3 &SNR     & \cite{Mori95} \cite{Rowell99}\\
PKS0521$-$322    &104.0 &AGN     & \cite{Roberts98} \\
PKS2005$-$489    & 94.5 &AGN     & \cite{Roberts99} \\
Cen A            & 80.3 &Radio galaxy & \cite{Susukita97} \cite{Rowell98}\\
PSR1055$-$52     & 78.6 &Pulsar  & \cite{Sako97} \cite{Susukita97}\\
SN1006           & 63.7 &SNR     & \cite{Tanimori98a}\\
RXJ1713.7$-$394  & 61.4 &SNR     & \cite{Muraishi99}\\
PKS0548$-$322    & 49.2 &AGN     & \cite{Roberts99} \\
PKS2155$-$304    & 37.5 &AGN     & \cite{Roberts99} \\
PKS2316$-$423    & 28.8 &AGN     & \cite{Roberts98} \\
Sgr A*           & 28.2 &Galactic center & \\
EXO0423.4$-$084  & 23.7 &AGN     & \cite{Roberts98} \\
GROJ1317$-$44    & 22.7 &Cen A? & \\
Vela X-1         & 20.8 &X-ray binary & \\
GRB970402        & 19.6 &Gamma-ray burst & \\
Cen X-3          & 17.7 &X-ray binary & \\
2EGJ1746-2852    & 15.2 &EGRET unID &  \\
\end{tabular}
\end{table}

\section*{Summary}

CANGAROO-III will start to explore
the southern half of the 100~GeV gamma-ray sky in 2004,
complementing projects located in the northern hemisphere
to ensure the entire sky is covered at these energies.

\end{document}